# Stochastic User Equilibrium Model with a Bounded Perceived Travel Time

Songyot Kitthamkesorn [a] and Anthony Chen [b]

[a] Excellence Center in Infrastructure Technology and Transportation Engineering, Department of Civil Engineering, School of Engineering, Chiang Mai University, 50200, Thailand

[b] Department of Civil and Environmental Engineering, The Hong Kong Polytechnic University, Hong Kong, P.R. China

**Abstract**:

Stochastic User Equilibrium (SUE) models depict the perception differences in traffic assignment problems. According to the assumption of an unbounded perceived travel time distribution, the conventional SUE problems result in a positive choice probability for all available routes, regardless of their unappealing travel time. This study provides an eUnit-SUE model to relax this assumption. The eUnit model is derived from a bounded probability distribution. This closed-form model aligns with an exponentiated random utility maximization (ERUM) paradigm with the exponentiated uniform distributed random error, where the lower and upper bounds endogenously determine the route usage. Specifically, a Beckmann-type mathematical programming formulation is presented for the eUnit-SUE problem. The equivalency and uniqueness properties are rigorously proven. Numerical examples reveal that the eUnit bound range between the lower and upper bounds greatly affects the SUE assignment results. A larger bound range increases not only the number of routes in the choice set but also the degree of dispersion in the assignment results due to a larger route-specific perception variance. The misperception is contingent upon the disparity between the shortest and longest travel times and the bounds. As the bound range decreases, the shortest route receives significant flow allocation, and the assignment result approaches the deterministic user equilibrium (DUE) flow pattern.

*Keywords*: Stochastic user equilibrium; bounded choice set; eUnit model; heterogeneous variance; mathematical programming

## 1. Introduction

### 1.1 Overview

Traffic flow forecasting is crucial for transportation planning. The deterministic user equilibrium (DUE) model (Wardrop, 1952) and the stochastic user equilibrium (SUE) model (Danzano and Sheffi, 1977) are two main methods for traffic assignment. The DUE model assumes perfect knowledge of transportation network conditions. All travelers are assumed to choose the shortest (or minimum cost) route for their journey. The SUE model relaxes the perfect knowledge assumption by introducing perceptual differences in alternate paths, where not all travelers end up selecting the minimum cost route.

The multinomial logit (MNL) model is the most employed route choice model that considers the imperfect knowledge of the transportation network conditions. Recent advancements in the MNL-SUE model have mostly centered around the relaxation of the assumption of independently and identically distributed (IID) with the Gumbel random error. Several sophisticated route choice models have been developed to account for the route overlapping problem (e.g., the C-logit model of Cascetta et al., 1996; the path-size logit (PSL) model of Ben-Akiva and Bierlaire, 1999; the cross-nested logit (CNL) model of Bekhor and Prashker, 1999; the paired combinatorial logit (PCL) model of Bekhor and Prashker, 1999) and/or the heterogeneous perception issue (e.g., Castillo et al., 2008; Kitthamkesorn and Chen, 2013; 2014; Nakayama, 2013; Nakayama and Makoto, 2015). The MNL-SUE model has an additional assumption that assigns a positive choice probability to all feasible routes in the choice set connecting an origin-destination (OD) pair, irrespective of their level of attractiveness. Indeed, it is worth noting that travelers encounter perception limitations when evaluating different routes (Bovy and Stern, 1990; Bovy, 2009). They avoid routes with high travel costs (Leurent, 1997; Prato and Bekhor, 2006; Prato, 2009; Watling et al., 2018). The process of route selection involves the exclusion of some routes based on individual restrictions and preferences, followed by the evaluation of a subset of available routes, ultimately leading to a final selection (Jan et al., 2000; Bekhor et al., 2006; Bovy, 2009; Gao et al., 2011; Kaplan and Prato, 2012). Limited study has been conducted to address such positivity assumption in the route choice set within the SUE framework.

### 1.2 Literature Review

Two SUE frameworks have been recently proposed in the literature to consider the perception limitation in determining routes in the choice set. The restricted SUE (RSUE) problem (Watling et al., 2015; Rasmussen et al., 2015; 2017) modified the MNL random utility maximization (RUM) model to include a predefined reference travel time across OD pairs as a bound. Routes having a travel time longer than the threshold are considered unappealing and are removed from the choice set. This results in a non-smooth choice probability at the threshold, and several undesired properties in the RSUE model, such as non-unique solution and computational intractability.

Watling et al. (2018) later introduced a bounded SUE (BSUE) model to overcome the drawbacks of the RSUE model. This was achieved by redefining the threshold. Instead of a direct modification of the MNL RUM deterministic term, a bound is applied to the difference between route travel time and the minimum travel time for creating a bounded choice (BC) model as follows:

$$P_r^{ij} = \frac{\left(exp\left(-\theta\left(g_r^{ij} - min\left(g_m^{ij}: m \in R_{ij}\right) - \rho^{ij}\right)\right) - 1\right)_+}{\sum_{k \in R_{ij}}\left(exp\left(-\theta\left(g_k^{ij} - min\left(g_m^{ij}: m \in R_{ij}\right) - \rho^{ij}\right)\right) - 1\right)_+}, \tag{1}$$

where $g_r^{ij}$ is the (mean) route travel time on route $r \in R_{ij}$ between OD pair $ij \in IJ$, $min\left(g_m^{ij}: m \in R_{ij}\right)$ is the minimum travel time between OD pair $ij \in IJ$, $\rho^{ij}$ is the threshold (or bound size) for the difference between the minimum travel time and the other routes' travel time of OD pair $ij$, $\theta$ is a scaling parameter, and $(x)_+ = max(x, 0)$. This closed-form model has the perceived travel time difference (i.e., $G_r^{ij} - min\left(G_m^{ij}: m \in R_{ij}\right)$) within the threshold as shown in Figure 1a. A fixed-point formulation for the BSUE model was presented. It can guarantee the uniqueness of the equilibrium solution of the used and unused routes from the continuous choice probability within the threshold. However, the non-smooth feature prevents the development of a convex mathematical programming (MP) formulation, which limits the adoption of well-developed solution algorithms for MP.

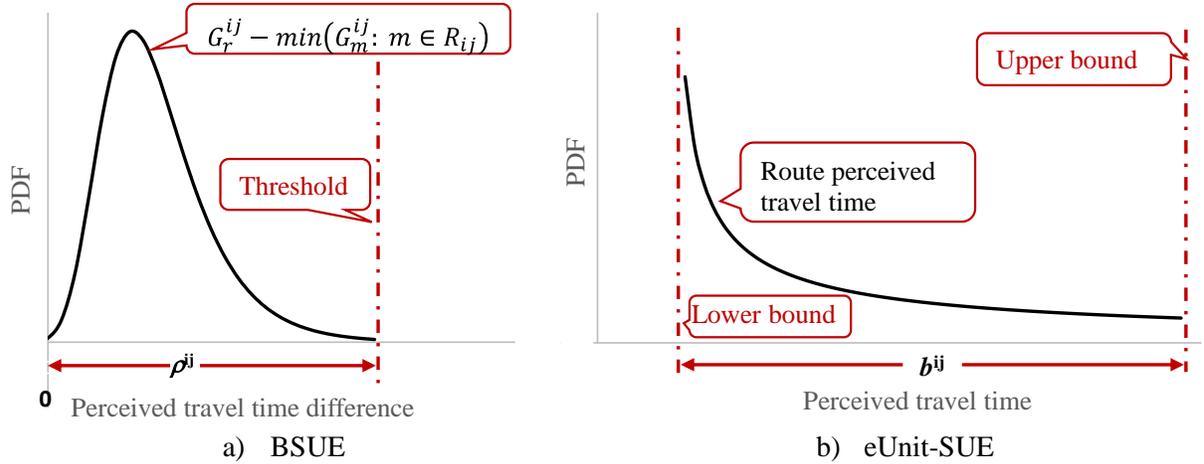

a) BSUE                                          b) eUnit-SUE

Figure 1: Perceived travel time considered in the BSUE and the proposed eUnit-SUE models

## 1.3 Objective and contributions

This study presents the development of a novel bounded route choice model and its corresponding convex MP formulation for the SUE problem. In contrast to modifying the Gumbel distribution, we adopt a bounded distribution, i.e., the exponentiated uniform distribution (Ramires et al., 2019), for the development of the eUnit route choice model as shown in Figure 1b. In this closed-form model, the bound range $b^{ij}$ between the upper and lower bounds between OD pair $ij \in IJ$ serves the purpose of not only determining the route usage, but also establishing the route-specific perception variance, which collectively exert a substantial influence on the choice probability. A Beckmann-type optimization model for the eUnit-SUE problem is developed using the convex MP formulation. The equivalency and uniqueness properties are rigorously proved. Numerical examples are also provided to demonstrate its features and applicability in large-scale transportation networks.

In summary, the main contributions of this paper are as follows:

1. The eUnit model is proposed to endogenously determine the route choice set from the upper and lower perceived travel time bounds.

2. A Beckmann-type MP formulation for the eUnit-SUE problem is developed.

The paper is organized as follows. The next section provides some background of the MNL and MNW models. Section 3 derives the eUnit model and its relationship with the existing models. An equivalent MP formulation of the eUnit-SUE problem is presented in Section 4. Section 5 shows some numerical examples, and Section 6 concludes the study.

## 2 Existing closed-form route choice models

Let $R_{ij}$ be a set of routes between origin−destination (OD) pair $ij \in IJ$ under consideration. This section provides a brief overview of some existing closed-form route choice models, including the multinomial logit (MNL) model and multinomial weibit (MNW) model.

### 2.1 Multinomial logit model

The MNL model is derived from the Gumbel distribution. The MNL random utility maximization (RUM) model can be written as an additive RUM or ARUM, i.e.,

$$U_r^{ij} = V_r^{ij} + \xi_r^{ij(L)}, \tag{2}$$

where $V_r^{ij}$ is the deterministic utility for traveling on route $r \in R_{ij}$ from between OD pair $ij \in IJ$, and $\xi_r^{ij(L)}$ is the Gumbel distributed random error under the independently and identically distributed (IID) assumption. By setting the (actual) travel time as the mean perceived travel time, the MNL RUM model can be written in terms of the travel disutility (travel cost or travel time) as (Dial, 1971)

$$U_r^{ij} = -\varrho^{(L)} g_r^{ij} + \xi_r^{ij(L)}, \tag{3}$$

where $g_r^{ij}$ is the (actual or mean) travel time for traveling on route $r$ from between OD pair $ij$, and $\varrho^{(L)}$ is the dispersion parameter. The MNL model has a closed-form choice probability expression, i.e.,

$$P_r^{ij} = \frac{exp\left(-\varrho^{(L)} g_r^{ij}\right)}{\sum_{k \in R_{ij}} exp\left(-\varrho^{(L)} g_k^{ij}\right)}. \tag{4}$$

According to the identically distributed assumption, the MNL model has homogeneous variance where all route choices have the same perception variance of $\pi^2/6\left[\varrho^{(L)}\right]^2$.

## 2.2 Multinomial weibit model

To relax the identically distributed assumption, Castillo et al. (2008) adopted the Weibull distribution to derive the multinomial weibit (MNW) model. The MNW RUM model can be expressed as a multiplicative utility or MRUM (Fosgerau and Bierlaire, 2009)

$$U_r^{ij} = \left(V_r^{ij}\right)^{-1} \xi_r^{ij(W)}, \tag{5}$$

where $\xi_r^{ij(W)}$ is the Weibull distributed random error with a unit shape parameter or the Exponential distributed random error. The MNW RUM model can be written in terms of the travel disutility as (Kitthamkesorn and Chen, 2013; 2014)

$$U_r^{ij} = \left(g_r^{ij} - \varsigma^{ij}\right)^{\varrho^{ij(W)}} \xi_r^{ij(W)}, \tag{6}$$

where $\varrho^{ij(W)}$ is the Weibull shape parameter, and $\varsigma^{ij}$ is the Weibull location parameter. Under the independently distributed assumption, the MNW can be written as

$$P_r^{ij} = \frac{\left(g_r^{ij} - \varsigma^{ij}\right)^{-\varrho^{ij(W)}}}{\sum_{k \in R_{ij}} \left(g_k^{ij} - \varsigma^{ij}\right)^{-\varrho^{ij(W)}}}. \tag{7}$$

Unlike the MNL model, the MNW model has a heterogeneous variance as a function of $g_r^{ij}$, i.e, (Kitthamkesorn and Chen, 2013; 2014; Gu et al., 2022)

$$\left(\sigma_r^{ij}\right)^2 = \left[\frac{\left(g_r^{ij} - \varsigma^{ij}\right)}{\Gamma\left(1 + \frac{1}{\varrho^{ij(W)}}\right)}\right]\left[\Gamma\left(1 + \frac{2}{\varrho^{ij(W)}}\right) - \Gamma^2\left(1 + \frac{1}{\varrho^{ij(W)}}\right)\right], \tag{8}$$

where $\Gamma(\ )$ is the Gamma function. The longer the travel cost, the larger the perception variance. According to the unbounded random error distribution, both MNL and MNW models generate a strictly positive choice probability on all possible routes, regardless of unappealing travel time.

## 3 eUnit model

This section develops the eUnit model from the Exponentiated Uniform distribution and shows that the eUnit model is consistent with an Exponetiated RUM (ERUM) model. The section begins with the Exponentiated Uniform distribution, followed by the model derivation, eUnit ERUM model, eUnit sensitivity analysis, and illustrative examples.

### 3.1 Exponentiated Uniform distribution

The exponentiated uniform distribution (Ramires et al., 2019) is an extension of the ordinary uniform distribution using the exponentiated class of distribution (Gupta and Kundu, 2001). The development purpose is to consider the increasing and bathtub feature from the hazard rate function. The exponentiated uniform cumulative distribution function (CDF) can be expressed as

$$F_r^{ij}(x) = \left(\frac{x - l_r^{ij}}{u_r^{ij} - l_r^{ij}}\right)^{\beta_r^{ij}}, \tag{9}$$

where $x \in \left(l_r^{ij}, u_r^{ij}\right)$, $u_r^{ij} > l_r^{ij} > 0$ are respectively the upper and lower bounds of the perceived travel time, and $\beta_r^{ij} > 0$ is the shape parameter. The expected value and the variance are as follows:

$$\mu_r^{ij} = \frac{u_r^{ij}\beta_r^{ij} + l_r^{ij}}{\beta_r^{ij} + 1},$$ (10)

$$\left(\sigma_r^{ij}\right)^2 = \frac{\beta_r^{ij}\left(u_r^{ij} - l_r^{ij}\right)^2}{\left(\beta_r^{ij} + 1\right)^2\left(\beta_r^{ij} + 2\right)}.$$ (11)

When $\beta_r^{ij} = 1$, the exponentiated uniform distribution collapses to the ordinary uniform distribution.

## 3.2 Model derivation

Under the independently distributed assumption, the choice probability of the exponentiated uniform distribution can be determined by

$$P_r^{ij} = \int_{l_r^{ij}}^{u_r^{ij}} H_r^{ij}(\ )dx,$$ (12)

where $H_r^{ij}(\ ) = \partial \prod_{k \in R_{ij}} F_k^{ij}(\ )/\partial x_r^{ij}$. Then, we have

$$P_r^{ij} = \int_{l_r^{ij}}^{u_r^{ij}} \frac{\beta_r^{ij}\left(\frac{x - l_r^{ij}}{u_r^{ij} - l_r^{ij}}\right)^{\beta_r^{ij}-1}}{\left(u_r^{ij} - l_r^{ij}\right)} \prod_{k \neq r}\left(\frac{x - l_k^{ij}}{u_k^{ij} - l_k^{ij}}\right)^{\beta_k^{ij}} dx.$$ (13)

Set $l_r^{ij} = l^{ij}$ and $u_r^{ij} = u^{ij}$ for all routes, we have

$$P_r^{ij} = \frac{\beta_r^{ij}}{\left(u^{ij} - l^{ij}\right)} \int_{l^{ij}}^{u^{ij}}\left(\frac{x - l^{ij}}{u^{ij} - l^{ij}}\right)^{\sum_{k \in R_{ij}}\beta_k^{ij}-1} dx,$$ (14)

which gives the choice probability, i.e.,

$$P_r^{ij} = \frac{\beta_r^{ij}}{\sum_{k \in R_{ij}}\beta_k^{ij}}.$$ (15)

## 3.3 eUnit ERUM model

The eUnit utility can be considered as the mean of the exponentiated uniform distribution in Eq. (10), i.e.,

$$V_r^{ij} = \frac{u^{ij}\beta_r^{ij} + l^{ij}}{\beta_r^{ij} + 1},$$ (16)

Substituting Eq. (16) into Eq. (15) gives the following closed-form probability:

$$P_r^{ij} = \frac{\frac{V_r^{ij} - l^{ij}}{u^{ij} - V_r^{ij}}}{\sum_{k \in R_{ij}}\frac{V_k^{ij} - l^{ij}}{u^{ij} - V_k^{ij}}}.$$ (17)

The eUnit utility compared with the MNL utility and MNW utility can be presented in Figure 2. The Gumbel probability density function (PDF) supports $(-\infty, \infty)$. The Weibull distribution has a lower bound of its location parameter $\varsigma^{ij}$. The exponentiated uniform distribution has both lower and upper bounds, i.e., $l^{ij}$ and $u^{ij}$. Then, the eUnit RUM model can be written as an exponentiated random utility maximization (ERUM) model, i.e.,

$$U_r^{ij} = \left(\varepsilon_r^{ij}\right)^{\left(\frac{V_r^{ij} - a^{ij}}{b^{ij} - V_r^{ij}}\right)^{-1}},$$ (18)

where $\varepsilon_r^{ij}$ is the exponentiated uniform distributed random error with $l^{ij} = 0$, $u^{ij} = 1$, and $\beta_r^{ij} = 1$ or the (unit) uniform distributed random error, i.e.,

$$F_{\varepsilon_r^{ij}}(x) = x.$$ (19)

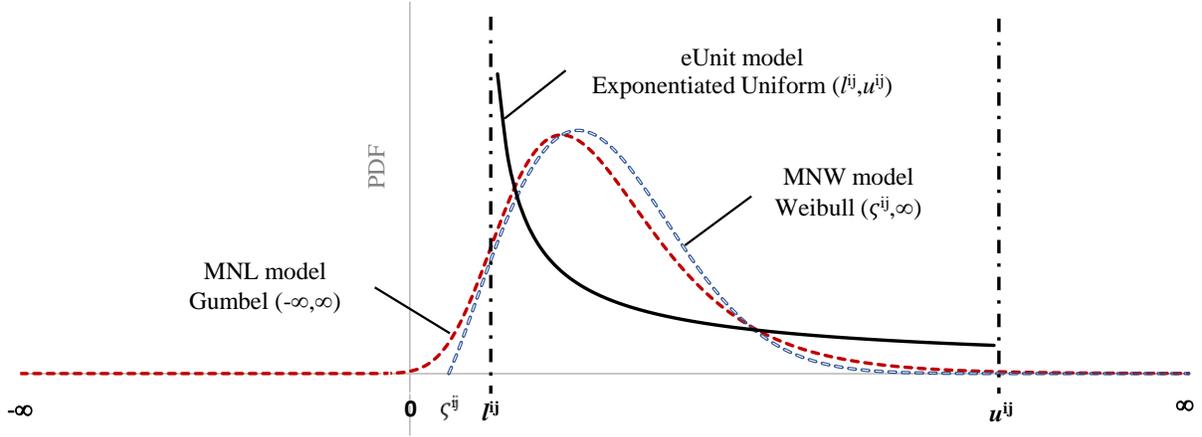

Figure 2: Utility PDF of the MNL model (Gumbel distribution), MNW model (Weibull distribution), and eUnit model (exponentiated uniform distribution)

The probability of choosing route $r$ can be determined by

$$P_r^{ij} = Pr\left( \left(\varepsilon_r^{ij}\right)^{\left(\frac{v_r^{ij}-l^{ij}}{u^{ij}-v_r^{ij}}\right)^{-1}} \geq \left(\varepsilon_k^{ij}\right)^{\left(\frac{v_k^{ij}-l^{ij}}{u^{ij}-v_k^{ij}}\right)^{-1}}, \forall k \neq r \right), \forall r \in R_{ij}, ij \in IJ,$$

$$P_r^{ij} = Pr\left( \left(\varepsilon_r^{ij}\right)^{\left(\frac{v_r^{ij}-l^{ij}}{u^{ij}-v_r^{ij}}\right)^{-1} \Big/ \left(\frac{v_k^{ij}-l^{ij}}{u^{ij}-v_k^{ij}}\right)^{-1}} \geq \varepsilon_k^{ij}, \forall k \neq r \right), \forall r \in R_{ij}, ij \in IJ. \tag{20}$$

Then, we have

$$P_r^{ij} = \int_0^1 \prod_{k \neq r} x_k \, dx_r. \tag{21}$$

From Eq. (20), we have

$$P_r^{ij} = \int_0^1 x^{\left(\frac{v_r^{ij}-l^{ij}}{u^{ij}-v_r^{ij}}\right)^{-1} \Sigma_{k \neq r} \left(\frac{v_k^{ij}-l^{ij}}{u^{ij}-v_k^{ij}}\right)} \, dx$$

$$P_r^{ij} = \left[ x^{\left(\frac{v_r^{ij}-l^{ij}}{u^{ij}-v_r^{ij}}\right)^{-1} \Sigma_{k \neq r} \left(\frac{v_k^{ij}-l^{ij}}{u^{ij}-v_k^{ij}}\right) + 1} \middle/ \left\{ \left(\frac{V_r^{ij}-l^{ij}}{u^{ij}-V_r^{ij}}\right)^{-1} \sum_{k \neq r} \left(\frac{V_k^{ij}-l^{ij}}{u^{ij}-V_k^{ij}}\right) + 1 \right\} \right]_0^1 \tag{22}$$

This gives the same choice probability as Eq. (17).

### 3.4 Relationship among ARUM, MRUM, and ERUM models

Some existing route choice models are a member of the ERUM model, e.g., the route choice model developed from the Pareto distribution in the semi-parametric approach (Li, 2011). The log transformation can link the ERUM model with the multiplicative RUM (MRUM) model (Fosgerau and Bierlaire, 2009) and the additive RUM (ARUM) model as shown in Figure 3. This is according to the distribution relations. The log Weibull distribution is the Gumbel distribution, and the log uniform distribution is the exponential distribution (Leemis and McQueston, 2008). The MNW model is equivalent to the MNL model through the logarithmic utility $-ln\,V_r^{ij}$ (Fosgerau and Bierlaire, 2009). When the eUnit model's lower bound equals the MNW location parameter, i.e., $l^{ij} = \varsigma^{ij}$, and the eUnit model's upper bound approaches infinity, the eUnit model is also related to the MNW model via the log transformation, i.e., $ln\left(V_r^{ij}-l^{ij}\right)$.

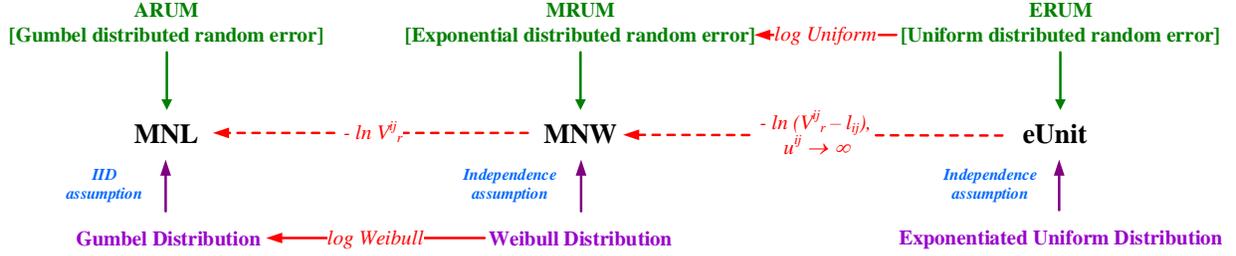

Figure 3: Relationship between the ARUM, MRUM, and ERUM models

From the above relationships, Eq. (18) is consistent with the MRUM model, i.e.,

$$U_r^{ij} = \left(\frac{V_r^{ij} - l^{ij}}{u^{ij} - V_r^{ij}}\right)^{-1} \ln \varepsilon_r^{ij},$$ (23)

where the logarithmic random error $\ln \varepsilon_r^{ij}$ is the exponential distributed random error. According to Fosgerau and Bierlaire (2009), $\left((V_r^{ij} - l^{ij})/(u^{ij} - V_r^{ij})\right)^{-1}$ can be considered as the eUnit (deterministic) utility. Let $g_r^{ij}$ be the travel time on route $r$ between OD pair $ij$. The eUnit disutility can be presented as $\left((u^{ij} - g_r^{ij})/(g_r^{ij} - l^{ij})\right)^{-1}$, and the eUnit RUM model can be expressed as

$$U_r^{ij} = \left(\varepsilon_r^{ij}\right)^{\left(\frac{u^{ij} - g_r^{ij}}{g_r^{ij} - l^{ij}}\right)^{-1}}.$$ (24)

Its corresponding choice probability can be expressed as

$$P_r^{ij} = \frac{\dfrac{u^{ij} - g_r^{ij}}{g_r^{ij} - l^{ij}}}{\sum_{k \in R_{ij}} \dfrac{u^{ij} - g_k^{ij}}{g_k^{ij} - l^{ij}}}.$$ (25)

As the Exponentiated Uniform distribution has bounds, $R_{ij}$ only includes the routes having the travel time within the specified upper and lower bounds $(l^{ij}, u^{ij})$.

**Definition 1.** eUnit choice probability

Since the eUnit model is based on a bounded perceived travel time distribution, the used route has the perceived travel time within $(l^{ij}, u^{ij})$. The eUnit choice probability can be expressed as

$$P_r^{ij} = \frac{\left(\dfrac{u^{ij} - g_r^{ij}}{g_r^{ij} - l^{ij}}\right)_+}{\sum_{k \in R_{ij}} \left(\dfrac{u^{ij} - g_k^{ij}}{g_k^{ij} - l^{ij}}\right)_+},$$ (26)

where $(x)_+ = max(x, 0)$.

### 3.5 Route-specific perception variance

From Eq. (11) and Eq. (16)ผิดพลาด! ไม่พบแหล่งการอ้างอิง, the eUnit route-specific perception variance can be expressed as a function of the travel time and the bounds, i.e.,

$$\sigma_{ijr}^2 = \frac{\dfrac{g_r^{ij} - l^{ij}}{u^{ij} - g_r^{ij}}\left(u^{ij} - l^{ij}\right)^2}{\left(\dfrac{g_r^{ij} - l^{ij}}{u^{ij} - g_r^{ij}} + 1\right)^2 \left(\dfrac{g_r^{ij} - l^{ij}}{u^{ij} - g_r^{ij}} + 2\right)}.$$ (27)

The route perception variance depends on the difference between the route travel time and the bounds as presented in Figure 4. The route perception variance is smaller near $l^{ij}$ or $u^{ij}$. The perception variance increases as the route travel time deviates from the bounds. The wider the bound range (i.e., $u^{ij} - l^{ij}$), the larger the perception variance.

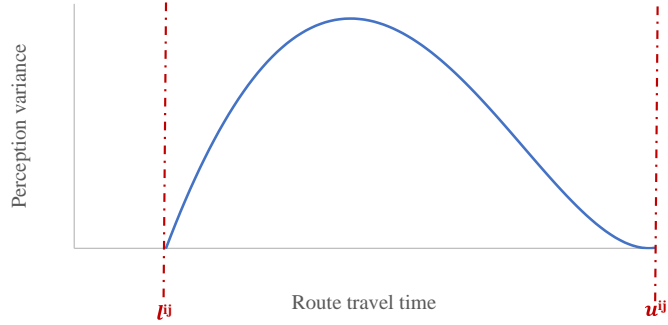

Figure 4: eUnit route perception variance

### 3.6 Sensitivity function

This subsection provides the sensitivity function of the eUnit model. According to Li (2011), this function represents a traveler's sensitivity to a change in travel time. The MNL model has the IID Gumbel perceived travel time. Its joint distribution when the scale parameter is fixed can be expressed as

$$\prod_{r \in R_{ij}} F_r^{ij}(x) = \prod_{r \in R_{ij}} exp\left\{-e^{-\varrho^{(L)}\left(x - \varphi_r^{ij}\right)}\right\} = \left[exp\left\{-e^{-\varrho^{(L)}x}\right\}\right]^{\sum_{r \in R_{ij}} exp\left(\varrho^{(L)}\varphi_r^{ij}\right)}, \quad (28)$$

where $\varphi_r^{ij}$ is the location parameter on route $r$ between OD pair $ij$. By relating $\varphi_r^{ij}$ to the (dis)utility (or route travel time), the MNL choice probability can be presented as a proportion of the exponent. Its sensitivity function can be expressed as

$$P_r^{ij} = \frac{exp\left(-\theta g_r^{ij}\right)}{\sum_{k \in R_{ij}} exp\left(-\theta g_k^{ij}\right)} = \frac{exp\left(S\left(g_r^{ij}\right)\right)}{\sum_{k \in R_{ij}} exp\left(S\left(g_k^{ij}\right)\right)}, \quad (29)$$

where $S(t)$ is the sensitivity function. Then, the MNL sensitivity function can be written as $S(t) = -\varrho^{(L)}t$, which indicates a linear sensitivity of travel time changes. According to the identical perception variance, the travelers are assumed to have an equal sensitivity to a unit change of the travel time under the MNL choice behavior. In contrast, the sensitivity function of the MNW model can be expressed as $S(t) = -\varrho^{ij(W)} ln \, t$. According to the logarithm function, travelers are more sensitive to an extreme end, the lower bound. It should be noted that Li (2011) assumed a value of zero for the Weibull location parameter. As $t > 0$ represents travel time, a unit change of travel time in a shorter route could impact more on the choice probability. This is consistent with the MNW route-specific perception variance where the shorter route has a smaller perception variance.

Following the same principle, the joint exponentiated uniform distribution under the independence assumption can be expressed as

$$\prod_{r \in R_{ij}} F_r^{ij}(x) = \left(\frac{x - l^{ij}}{u^{ij} - l^{ij}}\right)^{\sum_{r \in R_{ij}} \beta_r^{ij}}, \quad (30)$$

which gives the eUnit sensitivity function

$$S(t) = -ln\left(\frac{t - l^{ij}}{u^{ij} - t}\right). \quad (31)$$

From the eUnit sensitivity function, travelers are sensitive to both extreme ends (i.e., perceived travel time lower and upper bounds). As $t$ approaches the bound $l^{ij}$ or $u^{ij}$, the sensitivity function approaches $-\infty$ or $\infty$, respectively. A unit change of travel time in the shortest or longest route has a larger impact on the choice probability. The aforementioned extreme ends align with the eUnit route-specific perception variance described in Eq. (27), wherein the route with a travel time that is in close proximity to the bound exhibits a smaller perception variance.

### 3.7 Route choice probability

A three-route network in Figure 5 is adopted to present some characteristics of the eUnit model. In this network, there is a difference of 5 units in travel time between each route, and the middle route is the shortest path. For comparison purpose, the dispersion parameter of the MNL model and the scale parameter of the bounded choice (BC) model is set equal to 1. The shape and location parameters of the MNW model are assumed equal to 2.5 and 0, respectively (Kitthamkesorn and Chen, 2013; 2014). The BC model's difference threshold $\rho^{ij}$ equals 25. The eUnit model bound range is also 25 with $\left(l^{ij}, u^{ij}\right) = (4.75, 29.75)$. The MNL model cannot account for trip length due to the identically distributed assumption. The MNW model can take into account the route-specific perception variance, and the choice probability of each route gets increasingly similar as the trip length grows. The BC model yields the same result as the MNL model. This is due to the fact that the absolute difference between the shortest route and the other is the same for each x value. The eUnit model generates outcomes based on the value of x. We can divide the result into four cases according to the interaction between the travel time and the bounds as follows.

- *The value of x approaches zero.* The shortest travel time approaching $l^{ij}$ leads to a significant reduction of the route perception variance, resulting in a higher level of perception accuracy in the shortest route.
- *The value of x grows.* The difference between the shortest travel time and $l^{ij}$ increases. As a result, a larger route-specific perception variation makes the longer route more prone to deception.
- *The value of x increases, such that the longest route travel time converges to $u^{ij}$.* The difference between the longest travel time and $u^{ij}$ is decreasing. The perception of longer travel times is associated with a better level of accuracy, resulting in a decreasing probability of selecting the longest route.
- *The value of x increases, such that the travel time exceeds $u^{ij}$.* The route with a travel time greater than $u^{ij}$ is eliminated from the choice set.

In sum, the eUnit model has the ability to consider a bounded choice set similar to the BC model. Further, the proximity of the route travel time and the eUnit model's upper and lower bounds influence the choice probability. According to the eUnit model's route-specific perception variance, sensitivity function, and the above probability results, we define the following eUnit choice behavior.

**Definition 2.** eUnit choice behavior

Travelers possess lower and upper bounds to determine route usage. They further assume that the bound is progressively informative, and their decision-making process contingents upon the proximity of the route travel times in the choice set to both bounds.

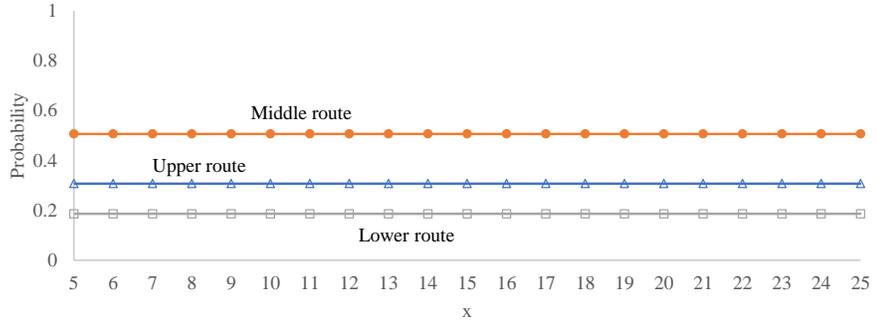

a)  MNL model

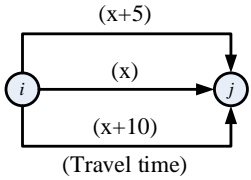

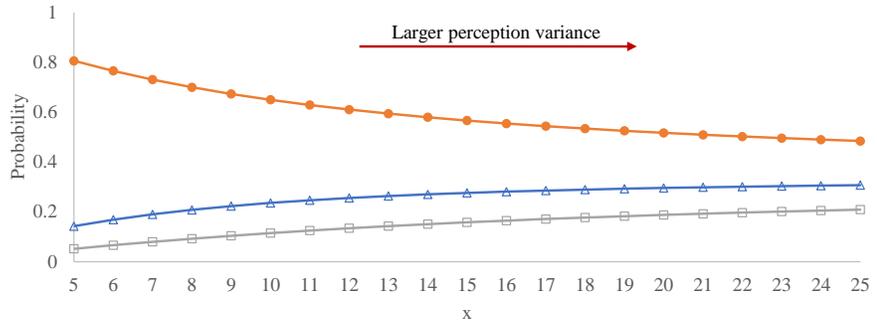

b)  MNW model

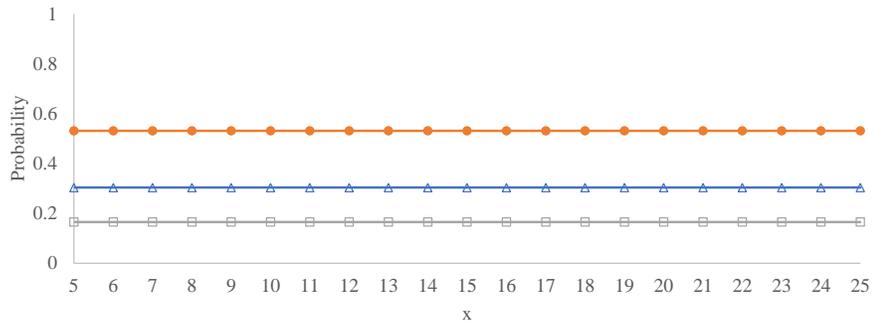

c)  BC model

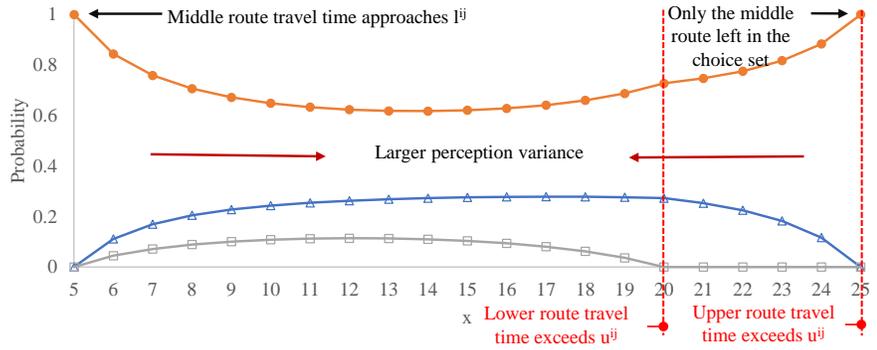

d)  eUnit model

Figure 5: Choice probability of the three-route network

# 4 Equivalent mathematical programming formulation

This section provides a mathematical formulation for the eUnit model in the stochastic user equilibrium (SUE) problem. The section begins with the eUnit-SUE mathematical programming formulation and is followed by the comparison between the eUnit-SUE and BSUE models.

## 4.1 Mathematical formulation

Consider the following mathematical programming (MP) formulation.

$$min \, Z = Z_1 + Z_2 = \sum_{a \in A} \int_0^{v_a} t_a(\omega) d\omega - \sum_{ij \in IJ} \sum_{r \in R_{ij}} b^{ij} ln(f_r^{ij} + 1), \quad (32)$$

s.t.

$$\sum_{r \in R_{ij}} f_r^{ij} = q_{ij}, \forall r \in R_{ij}, \quad (33)$$

$$f_r^{ij} \geq 0, \forall r \in R_{ij}, ij \in IJ, \quad (34)$$

where $f_r^{ij}$ is the flow on route $r$ between OD pair $ij$, $t_a$ is a strictly increasing travel time function w.r.t. its own traffic flow on link $a \in A$, $v_a = \sum_{ij \in IJ} \sum_{r \in R_{ij}} f_r^{ij} \delta_{ar}^{ij}$ is the flow on link $a$, $\delta_{ar}^{ij}$ equals 1 if link $a$ is on route $r$ and 0 otherwise, $b^{ij} = u^{ij} - l^{ij}$ is the bound range, and $q_{ij}$ is the OD travel demand. The objective function in Eq. (32) consists of Beckmann's transformation $Z_1$ and the logarithm term $Z_2$. Eq. (33) is the flow conservation constraint, and Eq. (34) is the non-negativity constraint. The term $Z_2$ and the conservation constraint in Eq. (33) generate the minimum perceived travel cost in the equivalency condition. It should be noted that $Z_2$ differs from the ordinary entropy term used in the MNL-SUE model (Fisk, 1980) and the MNW-SUE model (Kitthamkesorn and Chen, 2013), i.e., $f_r^{ij}[ln(f_r^{ij}) - 1]$. This logarithm term permits $f_r^{ij} = 0$ if the route is not in the choice set or the travel time is outside the two bounds.

**Remark 1.** At $b^{ij} = 0$, $Z_2$ is diminished. All routes are assumed to have zero perception variance, and the eUnit-SUE model collapses to the deterministic user equilibrium (DUE) model.

**Proposition 1.** The MP formulation presented in Eqs. (32)-(34) has the solution of the eUnit model.

**Proof.** See **Appendix A**.

**Proposition 2.** The solution of the eUnit-SUE model is unique.

**Proof.** See **Appendix B**.

**Remark 2.** Since the route flow is unique, the route travel time is unique. Accordingly, the route choice set generated by the eUnit-SUE model is also unique.

## 4.2 Comparison to the BSUE model

A two-route network is adopted to show the probability curve of the eUnit-SUE model and compared to the bounded SUE (BSUE) model in Figure 6. Watling et al. (2018) provided a fixed-point formulation to solve the BSUE condition. The BSUE model's assignment is based on the pairwise travel time difference and the threshold. Instead of a route travel time pairwise comparison, the eUnit-SUE model assumes that travelers have perceived travel time bounds, i.e., $l^{ij}$ and $u^{ij}$. As the route-specific perception variance of the eUnit model can be expressed by Eq. (27), the eUnit-SUE model's choice probability curve exhibits asymmetry. The curve is steeper as the route travel time approaches the bound, indicating a more informative to travelers consistent with the eUnit sensitivity function.

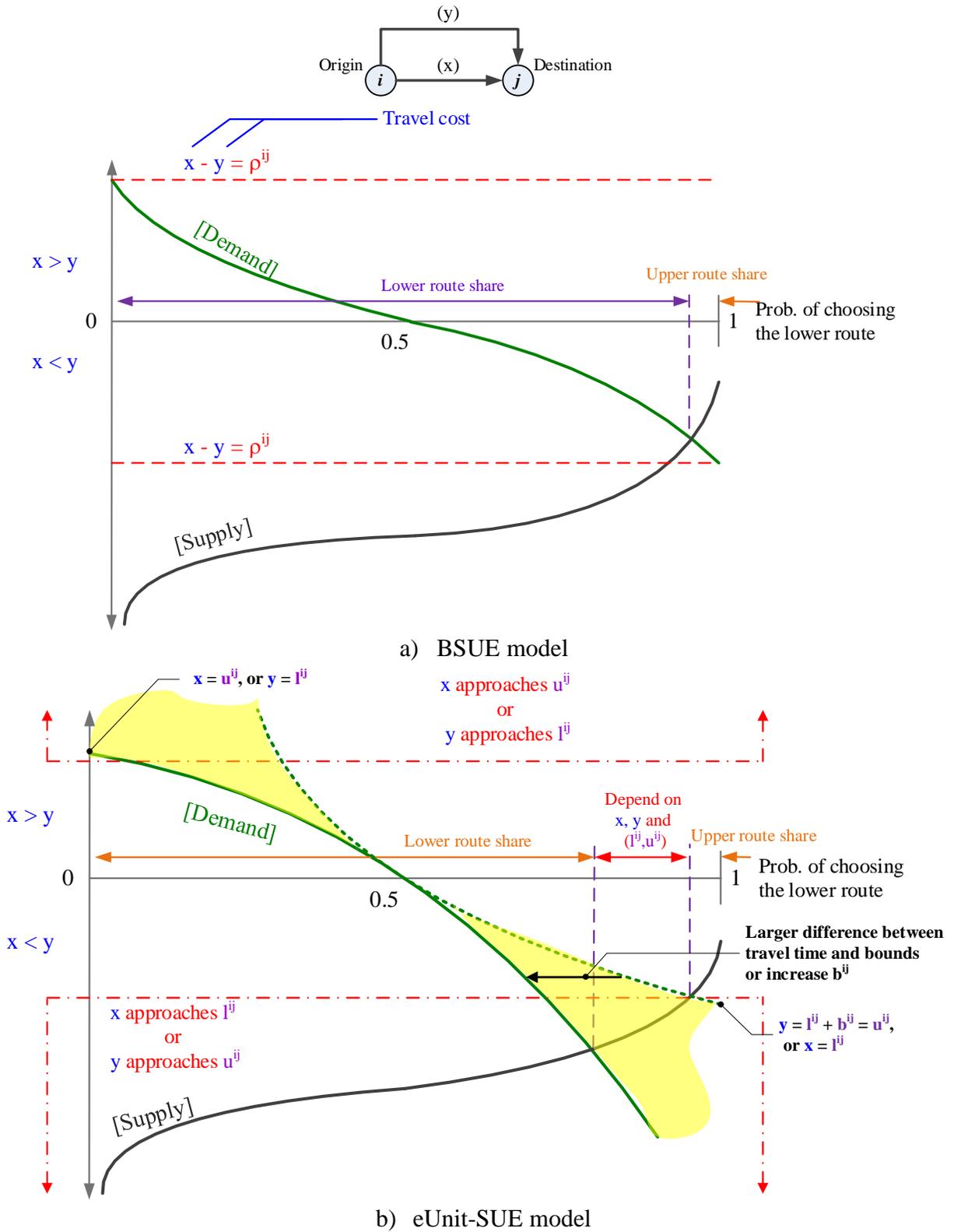

a) BSUE model

b) eUnit-SUE model

Figure 6: Visual illustration of the BSUE and eUnit-SUE models

## 5 Numerical examples

Two examples are given to investigate the characteristics of the eUnit-SUE model. A three-route network is used to demonstrate some effects of the bound range and features of the eUnit-SUE objective function. The Nguyen-Dupius network is employed to show the impacts of the congestion level under multiple OD pairs.

### 5.1 Three-route network

A three-route network shown in Figure 7 is adopted. This network has one OD pair with the travel demand of 100 units. Each route has the same capacity of 100 units. The travel time function parameters are varied due to different street classes. The scaling parameter of the BSUE model is 0.1, and the difference threshold $\rho^{ij} = 1$. The eUnit-SUE bound $b^{ij} = 1$. These settings are applied across the entire three-route network, unless otherwise specified.

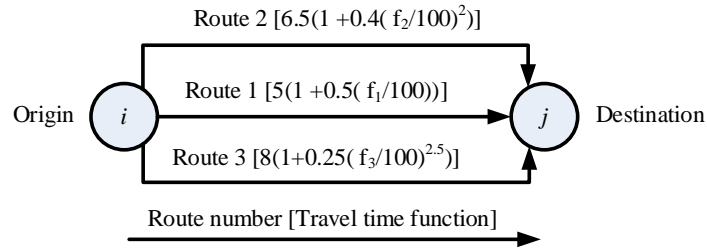

Figure 7: Three-route network

### 5.1.1 Equilibrium solution and travel time bound

Both BSUE and eUnit-SUE models exclude route 3 as depicted in Figure 8. The reason for this is that the travel time on route 3 exceeds the threshold ($u^{ij}$) of the BSUE (eUnit-SUE) model. Even though $\rho^{ij} = b^{ij}$, the resulting bound from these SUE models differs according to distinct choice behavior. The threshold of the BSUE model is 7.532, and the travel time on route 3 exceeds the threshold by 0.468. Given that $l^{ij} + b^{ij} = u^{ij}$, the eUnit-SUE model's perceived travel time lower bound is 6.715, and the perceived travel time upper bound is 7.715. The travel time on route 3 surpasses $u^{ij}$ by 0.285.

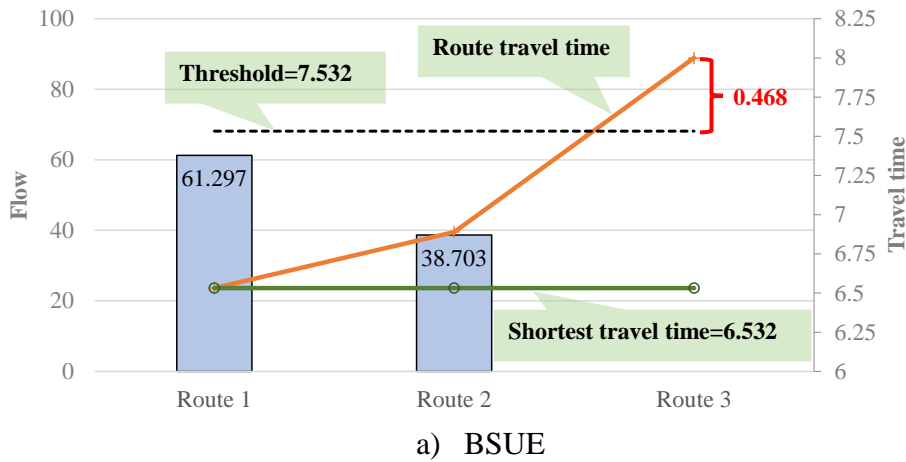

a) BSUE

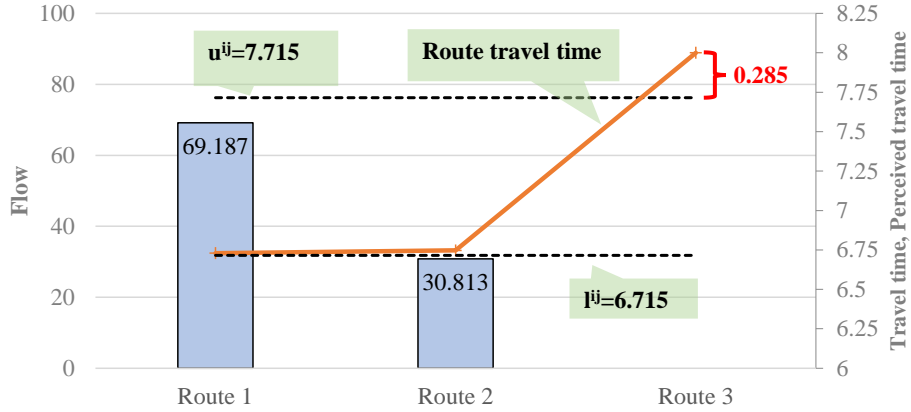

b) eUnit-SUE

Figure 8: Comparison between the BSUE and eUnit-SUE models at $\rho^{ij} = b^{ij} = 1$

### 5.1.2 Effect of bound range

The bound range plays an important role in the assignment result. When $\rho^{ij} = b^{ij} = 0$, the BSUE and eUnit-SUE models generate the result equivalently to the deterministic user equilibrium (DUE) model as presented in Figure 9. As $\rho^{ij}$ and $b^{ij}$ increase, both the BSUE and eUnit-SUE models exhibit a more comparable flow on each route and a greater disparity in travel time between the routes. The change in the bound appears to have a greater impact on the BSUE model. For the eUnit-SUE model, a larger $b^{ij}$ decreases (increases) $l^{ij}$ ($u^{ij}$) as presented in Figure 10. The resulting route flow difference stems from the heterogeneous perception variance. According to Eq. (27), the route-specific perception variance is larger as the route travel time gets distance from the bound. Consequently, the perception variance of the shortest path is higher. Travelers are assumed to increase their misperception of the travel time, and the travel demand is assigned more dispersedly.

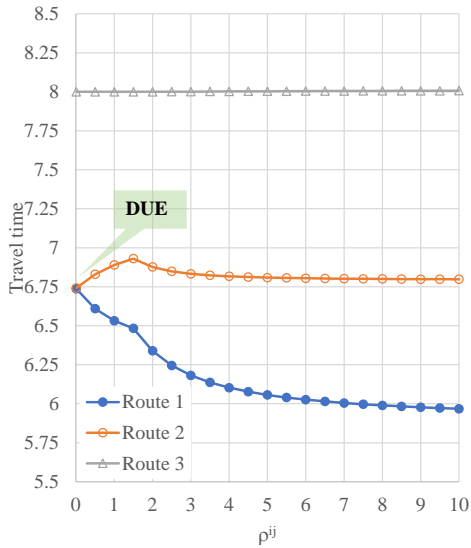

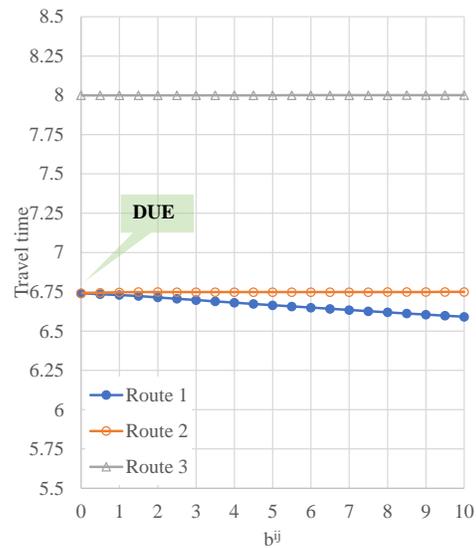

a) Travel time (BSUE)                              b) Travel time (eUnit-SUE)

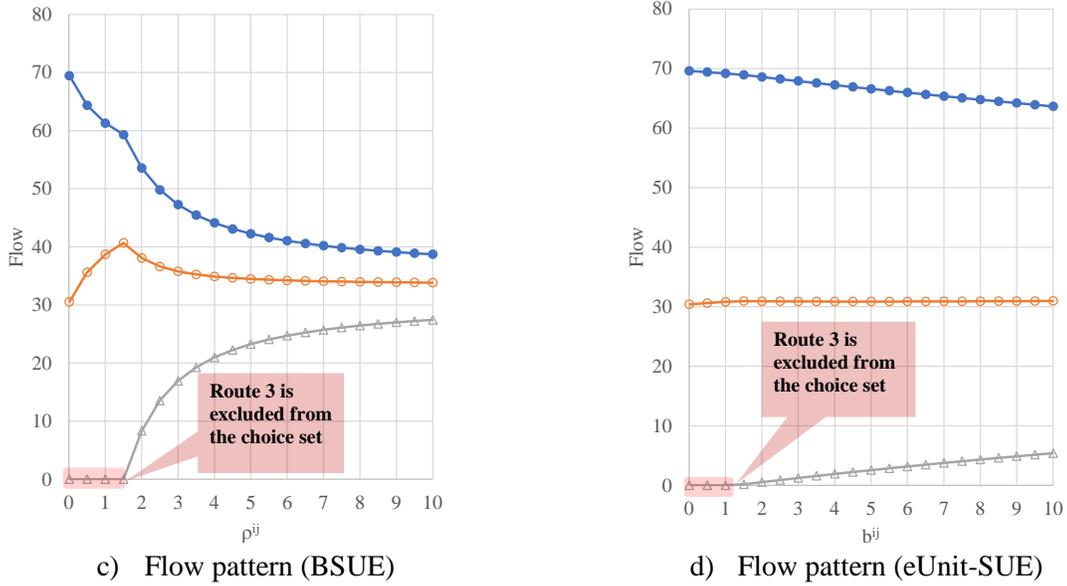

c)  Flow pattern (BSUE)                     d)  Flow pattern (eUnit-SUE)

Figure 9: Impact of bounds on the BSUE and eUnit-SUE model

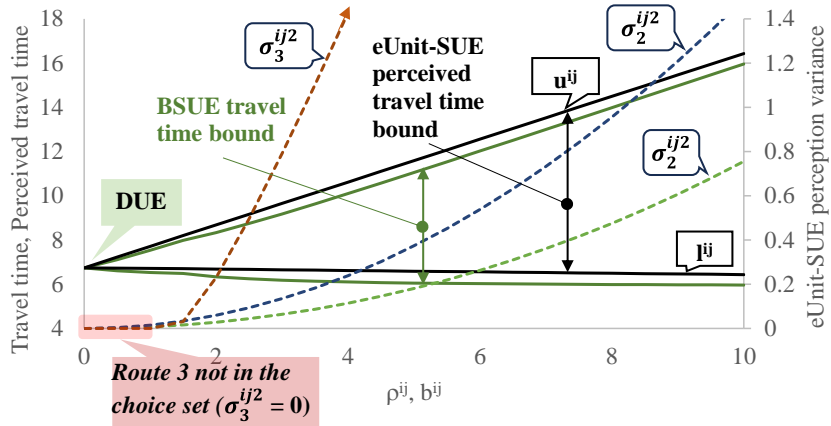

Figure 10: Impact of $\rho^{ij}$ and $b^{ij}$ on the bounds and perception variance

### 5.1.3 eUnit-SUE objective value

When $b^{ij}$ increases, the eUnit-SUE objective value decreases as presented in Figure 11. The increasing value of the logarithm term $Z_2$ is greater than the increasing value of Beckmann's term $Z_1$. Then, we consider the objective value at $b^{ij} = 1$ and $b^{ij} = 10$. When $b^{ij} = 1$, route 3 is omitted from the choice set. The zero flow under this SUE assignment is possible since the logarithm term $Z_2 = \sum_{ij \in IJ} \sum_{r \in R_{ij}} b^{ij} ln(f_r^{ij} + 1)$ allows $f_r^{ij} = 0$. This contrasts with the ordinary entropy term, i.e., $\sum_{ij \in IJ} \sum_{r \in R_{ij}} f_r^{ij} ln(f_r^{ij})$. The entropy term cannot allow zero flow according to $ln(0) \to \infty$. The assumption that $0 ln(0) = 0$ is required, and the flow assigned on a costly route can be expressed as a decimal value. At $b^{ij} = 10$, all three routes are included in the choice set and the flow on route 3 is positive.

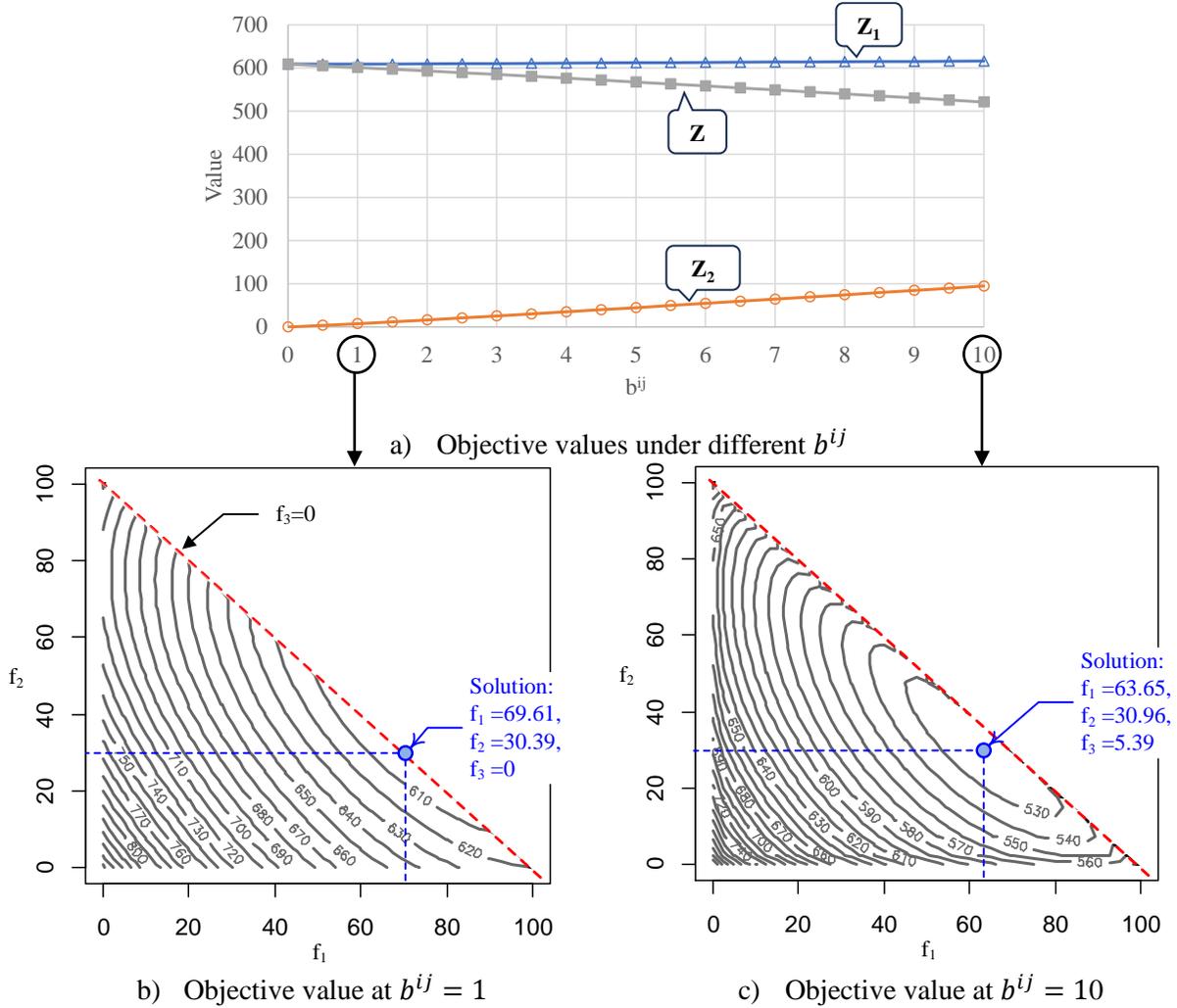

Figure 11: eUnit-SUE objective values

## 5.2 Nguyen-Dupius network

The Nguyen-Dupius network is adopted to show the impacts of multiple OD pairs and congestion level. The network topology is presented in Figure 12. This network has four OD pairs, i.e., (1, 2), (1, 3), (4, 2), and (4, 3). The total number of paths is 25. The link travel time follows the bureau of public road (BPR) function, i.e., $t_a = t_a^0(1 + 0.15(v_a/cap_a)^4)$, where $t_a^0$ is the free flow travel time (FFTT) on link $a$, and $cap_a$ is the capacity of link $a$. The scaling parameter of the BSUE model is 0.1. The difference threshold $\rho^{ij}$ and the eUnit-SUE bound $b^{ij}$ equal 10. Three scenarios of the demand level are investigated, including 50, 100, and 150 units for all OD pairs.

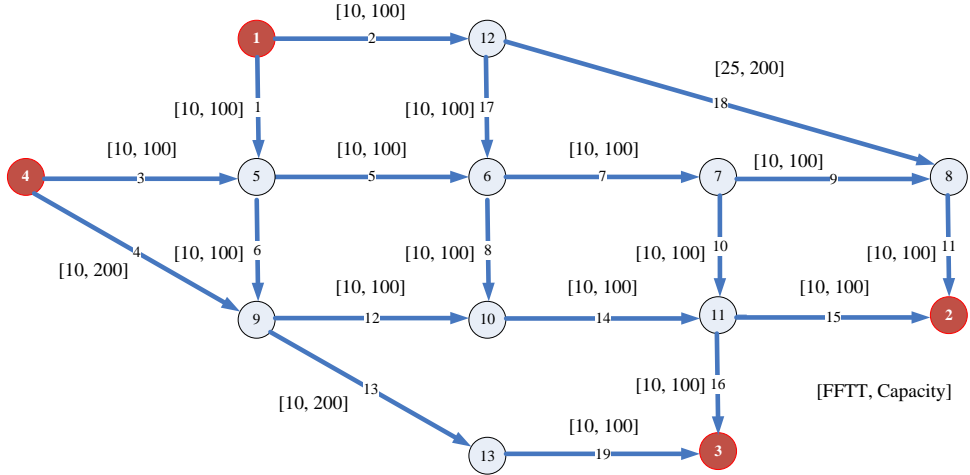

Route/Link

| OD | Route 1 | Route 2 | Route 3 | Route 4 | Route 5 | Route 6 | Route 7 | Route 8 |
|---|---|---|---|---|---|---|---|---|
| (1,2) | 1, 6, 12, 14, 15 | 1, 5, 8, 14, 15 | 1, 5, 7, 9, 11 | 1, 5, 7, 10, 15 | 2, 11, 18 | 2, 7, 9, 11, 17 | 2, 7, 10, 15, 17 | 2, 8, 14, 15, 17 |
| (4,2) | 1, 6, 13, 19 | 1, 6, 12, 14, 16 | 1, 5, 7, 10, 16 | 1, 5, 8, 14, 16 | 2, 8, 14, 16, 17 | 2, 7, 10, 16, 17 | | |
| (1,3) | 3, 5, 7, 9, 11 | 3, 5, 7, 10, 15 | 3, 5, 8, 14, 15 | 4, 12, 14, 15 | 3, 6, 12, 14, 15 | | | |
| (4,3) | 3, 5, 7, 10, 16 | 3, 5, 8, 14, 16 | 3, 6, 12, 14, 16 | 4, 13, 19 | 3, 6, 13, 19 | 4, 12, 14, 16 | | |

Figure 12: Nguyen-Dupius network

The bounds of both the BSUE and eUnit-SUE models exhibit an upward trend when the travel demand increases as shown in Figure 13. The BSUE model consistently generates similar bound patterns across all scenarios while the eUnit-SUE model yields diverse bound patterns for each scenario. The discrepancy in the number of routes within a choice set and the resulting assignment are attributed to the difference in choice behavior and, hence, the associated probability of route selection as presented in Table 1. Unlike the BSUE model, it should be noted that the eUnit-SUE model does not consistently yield the shortest travel time at the lower bound. In the eUnit-SUE model, there are two cases in which the shortest path exhibits a choice probability that is near to unity. The shortest travel time is in close proximity to $l^{ij}$, and the travel time of other routes approximate the value of $u^{ij}$. The majority of assignment results in this example adhere to the patterns observed in the first case. The second instance is observed in the OD pair $(4, 2)$ in scenario 2 and the OD pair $(4, 3)$ in scenario 3. These assignment results align with the eUnit choice behavior as described in **Definition 2** and illustrated in Figure 6. The perception variance is extremely small when the travel time near either $l^{ij}$ or $u^{ij}$. The travelers are assumed to unlikely be deceived by the travel time in this area.

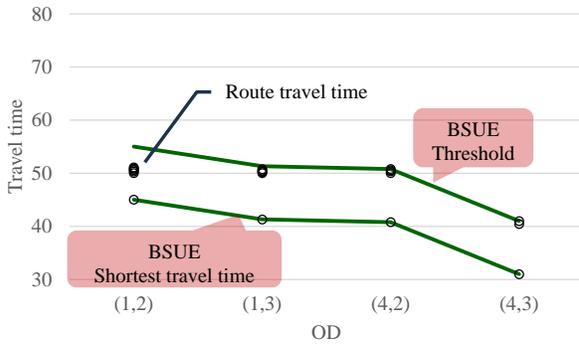

a) BSUE scenario 1

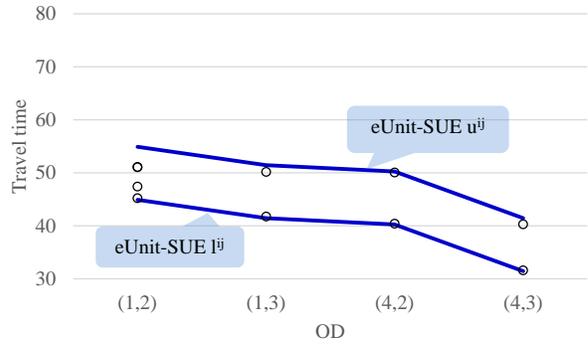

b) eUnit-SUE scenario 1

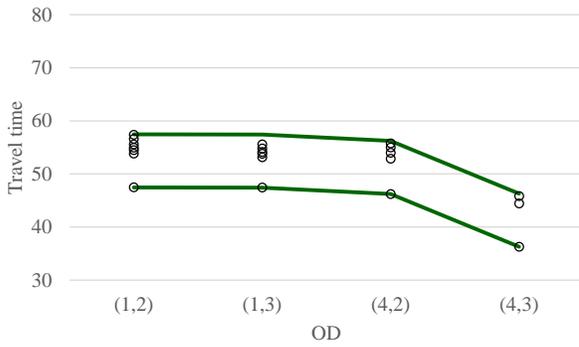

c) BSUE scenario 2

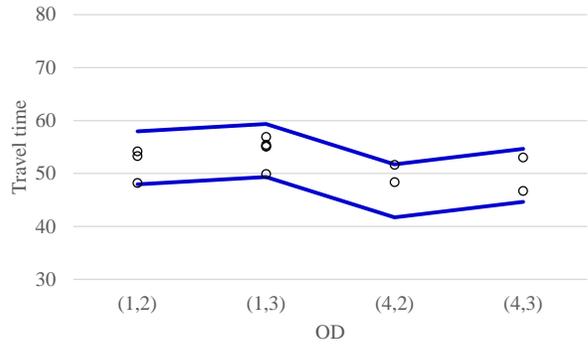

d) eUnit-SUE scenario 2

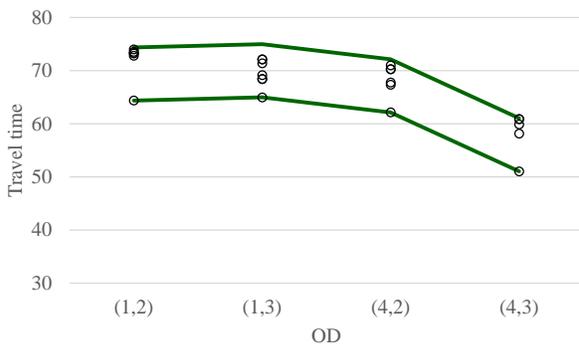

e) BSUE scenario 3

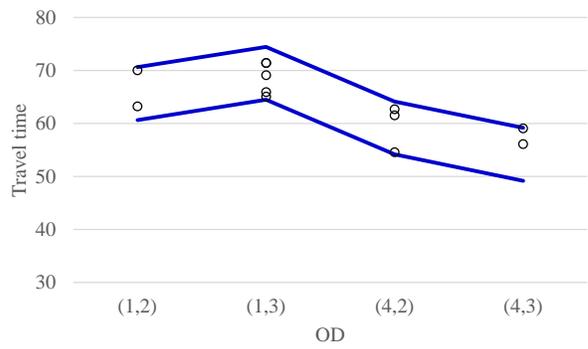

f) eUnit-SUE scenario 3

Figure 13: Bounds generated by BSUE and eUnit-SUE models for Nguyen-Dupius network

Table 1: Route choice probability under each scenario

| | | OD(1,2) | | | | | | | |
|---|---|---|---|---|---|---|---|---|---|
| Scenario | Model | Route 1 | Route 2 | Route 3 | Route 4 | Route 5 | Route 6 | Route 7 | Route 8 |
| 1 | BSUE | 19.93% | 10.57% | 11.67% | 10.86% | 31.97% | 0.00% | 12.08% | 0.00% |
| | eUnit-SUE | 0.00% | 0.00% | 1.79% | 1.76% | 87.84% | 0.00% | 8.61% | 0.00% |
| 2 | BSUE | 0.90% | 4.48% | 9.43% | 5.83% | 53.65% | 14.80% | 10.91% | 0.00% |
| | eUnit-SUE | 0.00% | 0.00% | 3.67% | 1.03% | 95.31% | 0.00% | 0.00% | 0.00% |
| 3 | BSUE | 0.00% | 0.00% | 4.49% | 6.90% | 81.17% | 2.59% | 4.84% | 0.00% |
| | eUnit-SUE | 0.00% | 0.00% | 0.00% | 1.21% | 98.79% | 0.00% | 0.00% | 0.00% |

| | | OD(4,2) | | | | |
|---|---|---|---|---|---|---|
| Scenario | Model | Route 1 | Route 2 | Route 3 | Route 4 | Route 5 |
| 1 | BSUE | 4.04% | 2.23% | 1.32% | 92.31% | 0.10% |
| | eUnit-SUE | 0.00% | 0.00% | 0.12% | 99.88% | 0.00% |
| 2 | BSUE | 14.16% | 9.14% | 6.59% | 67.90% | 2.21% |
| | eUnit-SUE | 0.00% | 0.00% | 0.08% | 99.92% | 0.00% |
| 3 | BSUE | 15.16% | 17.38% | 8.07% | 55.69% | 3.70% |
| | eUnit-SUE | 0.75% | 0.00% | 1.54% | 97.72% | 0.00% |

| | | OD(1,3) | | | | | |
|---|---|---|---|---|---|---|---|
| Scenario | Model | Route 1 | Route 2 | Route 3 | Route 4 | Route 5 | Route 6 |
| 1 | BSUE | 78.53% | 3.70% | 5.54% | 4.76% | 0.00% | 7.46% |
| | eUnit-SUE | 98.48% | 0.00% | 0.00% | 0.00% | 1.52% | 0.00% |
| 2 | BSUE | 48.61% | 7.85% | 13.30% | 11.64% | 0.00% | 18.60% |
| | eUnit-SUE | 83.58% | 1.59% | 4.23% | 3.22% | 3.69% | 3.68% |
| 3 | BSUE | 39.48% | 8.74% | 20.14% | 12.67% | 0.00% | 18.97% |
| | eUnit-SUE | 25.92% | 0.00% | 65.85% | 1.91% | 1.16% | 5.17% |

| | | OD(4,3) | | | | | |
|---|---|---|---|---|---|---|---|
| Scenario | Model | Route 1 | Route 2 | Route 3 | Route 4 | Route 5 | Route 6 |
| 1 | BSUE | 0.12% | 0.00% | 0.00% | 96.44% | 0.00% | 3.44% |
| | eUnit-SUE | 0.00% | 0.00% | 0.00% | 99.75% | 0.00% | 0.25% |
| 2 | BSUE | 0.62% | 0.00% | 0.00% | 86.55% | 1.70% | 11.13% |
| | eUnit-SUE | 0.00% | 0.00% | 0.00% | 88.17% | 0.00% | 1.83% |
| 3 | BSUE | 0.35% | 0.00% | 0.00% | 78.76% | 5.02% | 15.87% |
| | eUnit-SUE | 0.00% | 0.00% | 0.00% | 98.14% | 0.00% | 1.86% |

## 6. Conclusions and Suggestions

This study provided a bounded route choice model and its equivalent mathematical programming (MP) formulation for the stochastic user equilibrium (SUE) problem. The exponentiated uniform distribution (Ramires et al., 2019) was adopted to develop the eUnit model. This closed-form route choice model is consistent with the Exponentiated random utility maximization (ERUM) framework. The eUnit sensitivity function indicates that travelers are more extreme for a unit change in travel time at the bound, aligned with the eUnit route-specific perception variance as a function of the travel cost and the upper and lower bounds. A Beckmann's transformation-based MP formulation was presented for the eUnit-SUE problem, where a logarithm term was used to consider the flow-dependent bounds. The numerical examples revealed a significant impact of the demand level and bound range. The level of congestion influences the difference between the travel time and the bounds. The bound range has a significant impact on both the size of choice set and the choice probability results, especially for the shortest route. As the bound range reduces, the eUnit-SUE model's flow allocation approaches the deterministic user equilibrium traffic assignment.

Future research should explore some features of the ERUM model. Calibration of the parameters should be carried out. The convex MP formulation has the potential to simplify the application of the eUnit-SUE model to a realistic transportation network and allows further analyses of the eUnit-SUE model such as the sensitivity analysis of the equilibrium flow pattern. Similar to the existing models, consideration of the eUnit model in other choice dimensions in the SUE framework (e.g., Kitthamkesorn et al., 2016; 2017; Wang et al., 2020) and the location problem (Kitthamkesorn et al., 2021) could be possible.

### Acknowledgement

The work described in this paper was jointly supported by Chiang Mai University and the Hong Kong Polytechnic University.

## Appendix A. Proof of Proposition 1

The Karush-Kuhn-Tucker (KKT) condition can be presented as

$$g_r^{ij} - \frac{b^{ij}}{f_r^{ij} + 1} - \lambda_{ij} \geq 0, \tag{35}$$

$$f_r^{ij} \left( g_r^{ij} - \frac{b^{ij}}{f_r^{ij} + 1} - \lambda_{ij} \right) = 0, \tag{36}$$

where $\lambda_{ij}$ is the dual variable associated with the flow conservation constraint in Eq. (33). Assuming that $\lambda_{ij}$ is the minimum perceived travel cost between OD pair $ij$, i.e., $\lambda_{ij} = l^{ij}$, we have

$$f_r^{ij} \begin{cases} > 0 \; if \; g_r^{ij} - \dfrac{b^{ij}}{f_r^{ij}+1} - l^{ij} = 0 \\ = 0 \; if \; g_r^{ij} - \dfrac{b^{ij}}{f_r^{ij}+1} - l^{ij} \geq 0 \end{cases}. \tag{37}$$

The lower bound $l^{ij}$ is always less than each route travel cost $g_r^{ij}$ since $b^{ij} > 0$ and $f_r^{ij} \geq 0$. The flow $f_r^{ij}$ is greater than zero only if the route travel cost $g_r^{ij}$ is within $\left(l^{ij}, u^{ij}\right)$. Since $f_r^{ij}$ is non-negative as presented in Eq. (34), the operation $( \quad )_+$ in **Definition 1** is naturally satisfied, and no route violating the bound is being used.

Then, we have the route flow for $f_r^{ij} \geq 0$, i.e.,

$$f_r^{ij} = \frac{b^{ij} - g_r^{ij} + l^{ij}}{g_r^{ij} - l^{ij}}$$

$$f_r^{ij} = \frac{u^{ij} - g_r^{ij}}{g_r^{ij} - l^{ij}} \tag{38}$$

From Eq. (33) and Eq. (38), the OD travel demand can be presented as

$$q_{ij} = \sum_{r \in R_{ij}} f_r^{ij} = \sum_{r \in R_{ij}} \frac{u^{ij} - g_r^{ij}}{g_r^{ij} - l^{ij}}, \tag{39}$$

which leads to the eUnit route choice probability, i.e.,

$$P_r^{ij} = \frac{\dfrac{u^{ij} - g_r^{ij}}{g_r^{ij} - l^{ij}}}{\sum_{k \in R_{ij}} \dfrac{u^{ij} - g_k^{ij}}{g_k^{ij} - l^{ij}}}.$$

Thus, the MP formulation given in Eqs. (32)-(34) corresponds to the SUE model for which the route-flow solution $f_r^{ij} \geq 0$ is obtained according to the eUnit model. This completes the proof. □

**Appendix B. Proof of Proposition 2**

It is sufficient to prove that the objective function in Eq. (32) is strictly convex in the vicinity of route flow and that the feasible region is convex. The non-negative constraint has no effect on this property. This is accomplished by demonstrating that the Hessian matrix is positive definite. According to the link travel time is an increasing function w.r.t. its own flow, the Hessian matrix of Beckmann's transformation $Z_1$ is positive semidefinite w.r.t. the route flow variables. The Hessian matrix of $Z_2$ can be shown as

$$\frac{\partial^2 Z_2}{\partial f_r^{ij} \partial f_k^{ij}} = \begin{cases} \dfrac{b^{ij}}{\left(f_r^{ij}+1\right)^2}; \; r = k \\ 0 \; ; \; Otherwise \end{cases}. \tag{40}$$

Thus, the Hessian matrix of $Z_2$ is positive definite. Hence $Z = Z_1 + Z_2$ is strictly convex, and the eUnit-SUE solution is unique w.r.t. route flows. This completes the proof. □